\documentclass[aps,prb,twocolumn,showpacs,superscriptaddress,longbibliography,floatfix]{revtex4-2}  

\usepackage{graphicx}  
\usepackage{dcolumn}   
\usepackage{bm}        
\usepackage{amsmath}
\usepackage{amsthm}
\usepackage{amssymb}
\usepackage{bbold} 
\usepackage{mathrsfs}
\usepackage{array}
\usepackage[dvipsnames,usenames]{color}
\usepackage{soul} 
\usepackage{ulem} 
\usepackage[bookmarks=true,colorlinks,linkcolor=blue,urlcolor=blue,citecolor=blue]{hyperref}

\usepackage{physics}

\begin{document}

\normalem

\title{Single-site entanglement as a marker for quantum phase transitions at non-zero temperatures}

\author{Willdauany C. de Freitas Silva}
\affiliation{Instituto de Física, Universidade Federal do Rio de Janeiro, Rio de Janeiro, RJ, 21941-972, Brazil}

\author{Andressa R.~Medeiros-Silva}
\affiliation{Instituto de Física, Universidade Federal do Rio de Janeiro, Rio de Janeiro, RJ, 21941-972, Brazil}
\affiliation{Department of Physics, University of Houston, Houston, Texas, 77004, USA}

\author{Rubem Mondaini}
\affiliation{Department of Physics, University of Houston, Houston, Texas, 77004, USA}
\affiliation{Texas Center for Superconductivity, University of Houston, Houston, Texas, 77204, USA}

\author{Vivian V. França}
\affiliation{Instituto de Química, São Paulo State University, 14800-090, Araraquara, São Paulo, Brazil}

\author{Thereza Paiva} 
\affiliation{Instituto de Física, Universidade Federal do Rio de Janeiro, Rio de Janeiro, RJ, 21941-972, Brazil}

\begin{abstract}
Entanglement has been widely investigated in condensed matter systems since they are considered good candidates for developing quantum technologies. Additionally, entanglement is a powerful tool to explore quantum phase transitions in strongly correlated systems, with the von Neumann entropy being considered a proper measure of quantum entanglement for pure bipartite systems. For lattice systems, in particular, the single-site entanglement quantifies how much information about the quantum state of the remaining sites can be obtained by a measurement at a single site. Here, we use Quantum Monte Carlo calculations to obtain the average single-site entanglement for the two-dimensional Hubbard model in different geometries, probing the effects of varying temperature and interaction strength. We find that the average single-site entanglement signals the quantum phase transitions in such systems, allowing us to identify and characterize signatures of quantum phase transitions even at finite temperatures. We also analyze the relation between entanglement and magnetic susceptibility: in all the geometries considered, we find regimes in which the quantities are linearly connected. Our findings could then guide experiments to estimate entanglement via the susceptibility.
\end{abstract}

\maketitle

\section{Introduction}

Entanglement, a key property of quantum mechanics, is considered a fundamental asset for quantum information processes such as computation and communication, thus crucial for the development of quantum technologies \cite{Vedral_2003,klich_2006,Acin_2018,Wei_2020}. Condensed matter systems are particularly promising for quantum technology devices due to their inherent quantum fluctuations \cite{amico_2008,Islam_2015,Aslam_2023,Hempel_2018}. In correlated many-body and cold-atoms systems, entanglement also plays an important role in detecting and characterizing quantum phases of matter \cite{Vidal_2003,Osterloh_2002, Zanardi_2002, Gu_2004,Brukner_2006,Franca_2007,chakraborty_2012,Ghosh_2003}. 

For the particular case of the one-dimensional (1D) Hubbard model $-$ one of the primary models used to describe strongly interacting fermions $-$ bipartite entanglement has been used to explore quantum phase transitions at $T=0$, such as metal-insulator transitions\;\cite{Larson_2006, Deng_2006, Brunner_2013,Mendes_2013,Canella_2021,Canella_2022} and superfluid-insulator transitions\;\cite{Anfossi_2006}, including in disordered systems\;\cite{Canella_2019,Zawadzki_2024} and conventional superfluid to exotic superfluid transitions in spin-imbalanced systems\;\cite{Franca_2017,Arisa_2020}.
The restriction to zero temperature arises from the von Neumann entropy, which is a proper measure of bipartite entanglement only for pure states.

In contrast, in two-dimensional fermionic systems, entanglement has been less explored since numerical techniques often employed on 1D systems start facing limitations. Going beyond the computation of free-fermionic systems in either Hermitian\;\cite{Hyejin_2012} or non-Hermitian settings\;\cite{Yi_2021}, the work by Grover\;\cite{Grover2013} pioneered a method to study the Rényi entanglement entropy (REE) in auxiliary-field quantum Monte Carlo methods\;\cite{blankenbecler_1981, Hirsch_1985}. Challenges associated with the statistical convergence of the estimator in the importance sampling in the simulations gave rise to improved approaches\;\cite{Broecker_2014, Assaad_2015}, and for the search of additional schemes where one could compute it reliably\;\cite{Gaopei_2023, Zhang_2024}. For instance, it is now possible to establish subleading corrections to the area law in the REE within determinant quantum Monte Carlo (DQMC) for interacting fermions in the honeycomb lattice\;\cite{Jonathan_2024}.  

Another important feature of entanglement is that it has been associated with the magnetic susceptibility\;\cite{souza_2008, das_2013, Larson_2005, Arisa_2020}, a thermodynamic observable that could thus be used for estimating entanglement experimentally, as in spin chains systems formed by the compounds Na$_{2}$CU$_{5}$Si$_{4}$O$_{14}$\;\cite{souza_2008} and Cu(NO$_{3})_{2} \times 2.5 $H$_{2}$O\;\cite{das_2013}. In particular, studies in homogeneous one-dimensional Hubbard chains at zero temperature have shown that the single-site entanglement entropy is directly connected to the susceptibility\;\cite{Larson_2005} and that there are regimes at which this relation is linear and universal\;\cite{Arisa_2020}, suggesting that entanglement could be estimated experimentally via magnetic susceptibility measurements.

Here, we investigate whether the von Neumann entropy at $T\neq0 $ can still define bipartite entanglement and/or be used to identify and characterize Mott metal-insulator transitions (MITs) in two-dimensional lattices. That is, if signatures of the $T=0$ quantum phase transition can be recognized at finite temperatures, while also analyzing the connection between entanglement and magnetic susceptibility at $T\neq 0$. We consider the Hubbard model on honeycomb and kagome lattices, and the ionic Hubbard model on the square lattice, using DQMC to calculate the single-site entanglement at half-filling in all cases. Our results reveal that it is possible to recover the transition points previously encountered in the literature for all the lattices geometries and demonstrate that the entanglement entropy serves as a reliable indicator of quantum phase transitions to a Mott state, even at finite temperatures ($T \neq 0$). Additionally, for all the geometries studied, we find regimes at which entanglement and susceptibility are linearly related: this could then be used to estimate entanglement in experiments via susceptibility measurements.

The presentation is organized as follows: We introduce the model, the method, and how we compute the single-site entanglement entropy, added by a brief description of the systems of interest, in Sec.~\ref{section2}; in Sec.~\ref{Results} we present our results and discussions; finally, Sec.~\ref{conclusions} summarizes our findings. 

\section{Model and Methodology} \label{section2}

\subsection{Hamiltonian}

\begin{figure}[t]
\includegraphics[width=0.98\linewidth]{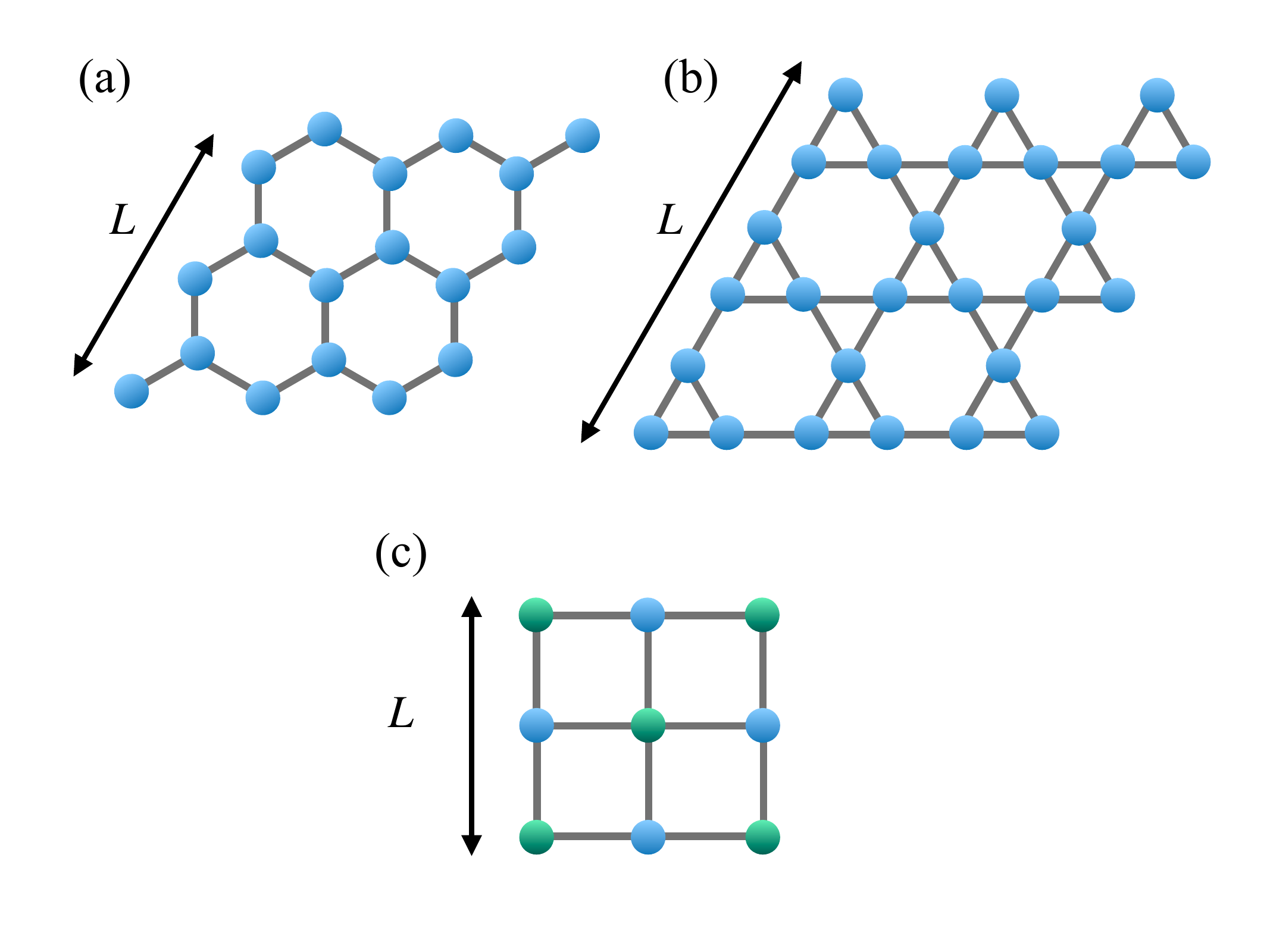}  
\caption{Schematic representation of the honeycomb (a), kagome (b), and square (c) lattices used, with $N = 2L^2, 3L^2$ and $L^2$ sites, respectively ($L = 3$ here). The latter exhibits a breaking of sublattice symmetry associated with the ionic Hubbard model, identified by the different colors.}
\label{fig:lattices}
\end{figure}

The Hubbard model Hamiltonian is given by
\begin{align}\label{Eq:dqmc_hamil}
\nonumber \hat {\cal H} = & -t\sum_{\substack{\langle \textbf{i},\textbf{j} \rangle},\sigma} \big( \hat c_{\textbf{i} \sigma}^{\dagger}\hat c_{\textbf{j} \sigma}^{\phantom{\dagger}}+ {\rm H.c.} \big) - \mu \sum_{\substack{\textbf{i}}, \sigma} \hat n_{\textbf{i},\sigma}
\\  & + U   \sum_{\substack{\textbf{i}}} \big(\hat n_{\textbf{i},\uparrow} - 1/2 \big) \big(\hat n_{\textbf{i},\downarrow} - 1/2\big),
\end{align}
where the sums run over all $N$ sites of the lattice. The second-quantized operators $\hat c^{\dagger}_{\mathbf{i} \sigma}$ ($\hat c^{\phantom{\dagger}}_{\mathbf{i} \sigma}$) describe the creation (annihilation) of electrons on a site $\mathbf{i}$, with spin $\sigma$, while $\hat n_{\mathbf{i}\sigma} \equiv \hat c^{\dagger}_{\mathbf{i} \sigma} \hat c_{\mathbf{i} \sigma}^{\phantom{\dagger}}$ are the corresponding number operators. In Eq.\,\eqref{Eq:dqmc_hamil}, the first two terms on the right-hand side describe the fermionic itinerancy between nearest-neighbor sites $\langle \mathbf{i}, \mathbf{j} \rangle$, and the filling of the bands, governed by the chemical potential $\mu$. The third term gives the local (onsite) repulsive interaction to which they are subject, with strength $U$. Henceforth, we define the hopping integral $t$ as the energy scale and the lattice constant as unity. We here study the Hubbard model on the honeycomb and kagome lattices shown in Figs.\;\ref{fig:lattices}(a) and \ref{fig:lattices}(b), respectively.

The ionic Hubbard model on a square lattice is also studied; its Hamiltonian reads:

\begin{align}\label{Eq:ihm_hamil}
\hat {\cal H}_{\rm ionic} = \hat {\cal H} + \Delta \sum_{\substack{\textbf{i}}} (-1)^{\textbf{i}} \hat n_{\bf i}\ ,
\end{align}
which differs from Eq.\;\eqref{Eq:dqmc_hamil} by the addition of a term that differentiates the two sublattices of the square lattice [shown in green and blue in Fig.\;\ref{fig:lattices}(c)], by adding (subtracting) an onsite energy $\Delta$ to even (odd) sites.

\subsection{Determinant quantum Monte Carlo}
\label{DQMC_method}

To investigate the Hamiltonians discussed, we employ the determinant quantum Monte Carlo method \;\cite{blankenbecler_1981,hirsch_1983,white_1989, Becca_2017}. This method maps a $d$-dimensional system of interacting fermions onto a $d+1$-dimensional system of non-interacting fermions coupled to an auxiliary field. To achieve this, two steps are performed. The first is to employ a Suzuki-Trotter decomposition in the partition function to separate the exponentials of the single-body terms from the two-body terms, thereby increasing the system's dimension. The additional dimension maps the imaginary time $L_{\tau}$ as a discretization of the inverse temperature $\beta = 1/T$ ($k_B=1$), such that $L_\tau \Delta \tau=\beta$. This procedure introduces a controllable error proportional to $(\Delta \tau)^2$, being exact as $\Delta \tau \rightarrow 0$. We set $t\Delta \tau \leq 0.1$  to ensure that the statistical errors from the Monte Carlo method are larger than the Trotter errors.

The second step is to reduce the order of the quartic operators using a Hubbard-Stratonovich transformation\;\cite{hirsch_1983}, turning them into quadratic single-particle operators by introducing bosonic auxiliary fields, which are then sampled using Monte Carlo techniques\;\cite{Santos_2003,assaad02,Gubernatis_2016,Becca_2017}. The correlation functions of interest can then be computed from matrix elements of Green's functions, a central object in the calculations, via the Wick theorem.

Nonetheless, the Monte Carlo weight can become negative for certain configurations of the auxiliary field, such as in systems that lack particle-hole symmetry (PHS). This phenomenon is known as the sign problem\;\cite{Santos_2003, mondaini2022, gaopei_2024}. To calculate the quantities of interest, one discards the negative sign from the Monte Carlo weight, effectively reweighting the average values in which any observable is scaled by the inverse of the average sign,  $\langle {\rm sign} \rangle$. This introduces large fluctuations when studying regimes with smaller values of $\langle {\rm sign} \rangle$, such as lower temperatures, higher values of $U$, and in larger systems\,\cite{white_1989, Loh_1990}. To mitigate the resulting fluctuations, acquiring a statistically relevant signal-to-noise ratio requires running simulations with an exponentially large number of Monte Carlo steps\;\cite{Hirsch_1985,Loh_1990,Troyer_2005}. 

\subsection{Single-site entanglement}
\label{singlesite}


Among the various entanglement measures\;\cite{Pauletti_2024}, the von Neumann entropy\;\cite{Neumann_1955}, part of the general class of Rényi entropies\;\cite{renyi_1961, Jose_2010}, stands out as an efficient tool to quantify the entanglement of bipartite pure states. One particular approach to the von Neumann entropy is the \textit{single-site} entanglement,  which has been extensively explored in the one-dimensional Hubbard model\;\cite{Larson_2005, Zanardi_2002, Gu_2004,wu_2004, Anfossi_2006, Franca_2008,capelle_2013, Franca_2017,picoli_2018,ferreira_2022}. In this case, the calculated entanglement is between a single site and the rest of the system. This measure is particularly relevant when it is possible to control and detect single particles in quantum systems, such as in cold atoms\;\cite{blatt_2012,bloch_2012}. For a single site $i$, the entanglement with the remaining $N - 1$ sites is then\;\cite{Canella_2019,Arisa_2020}

\begin{equation}
\label{entropy}
S_i = -\frac{1}{\log_2 d}\mathrm{Tr} [\rho_i \log_2 \rho_i]\ ,
\end{equation}
where $\rho_i = \mathrm{Tr}_{\displaystyle\substack{\{N - i\}}} \rho$ is the reduced density matrix, $\rho$ is the ground state's total density matrix\;\cite{Zanardi_2002}, $d$ is the Hilbert space dimension and $1/\log_2d$ is the normalization factor. As we are considering a single site, the local Hilbert space has $d=4$, considering the four possibilities $\ket{\uparrow}$, $\ket{\downarrow}$, $\ket{\uparrow \downarrow}$, and $\ket{0}$. In particular, $S_i$ can be rewritten in terms of the occupation probabilities\;\cite{Gu_2004, Larson_2005, Franca_2006,Franca_2017} as
\begin{multline}
 S_i = {-\frac{1}{\log_2 d}}[\omega_{i,\uparrow}\log_2 \omega_{i,\uparrow} +\omega_{i,\downarrow}\log_2 \omega_{i,\downarrow} \\
+\omega_{i,\uparrow \downarrow}\log_2 \omega_{i,\uparrow \downarrow}  +\omega_{i,0}\log_2 \omega_{i,0}] \nonumber,
\end{multline}
where the single occupation probability, owing to the SU(2) symmetry of the studied models, is the same for spin up and spin down and is given by $\omega_{i,\uparrow} = \omega_{i,\downarrow} = \langle \hat n_i\rangle/2 -  \omega_{i,\uparrow \downarrow}$, where $\omega_{i,\uparrow \downarrow} = \langle \hat{n}_{i,\uparrow} \hat n_{i,\downarrow}\rangle$ is the double occupancy probability at site $i$, and $\omega_{i,0}=1-\langle \hat n_i\rangle+\omega_{i,\uparrow \downarrow}$ is the empty occupation probability.

Notice that $S_i$ is zero whenever one of the probabilities is one (with the remaining being zero), i.e., when the reduced system is in a pure state.
For the Hubbard model, this occurs in the vacuum state $n \equiv \sum_i \langle \hat n_i\rangle=0$ (thus $\omega_{i,0}=1$) or at full-filling $n =2$ ($\omega_{i,\uparrow\downarrow}=1$). The maximum entanglement, $S_i=1$, occurs for the maximally mixed reduced system. This is the case when all four probabilities are equivalent, $\omega_i=1/4$, which occurs at $n=1$ and $U=0$\;\cite{Franca_tese_2008}. At half-filling, in the presence of repulsive interactions ($U>0$), the single occupation probabilities reach $\omega_\uparrow=\omega_\downarrow=0.5$ as $ U\to\infty$, and the double and empty probabilities tend to zero, leading to $S_i=0.5$. This remaining entanglement is then attributed to spin degrees of freedom.

To improve statistics in our calculations, we consider the average over all sites $S \equiv \frac{1}{N}\sum_i^NS_i$,\;\cite{Arisa_2020, Pauletti_2024} for the honeycomb and kagome lattices, thus obtaining a global measure representing the whole system. For the ionic Hubbard model on a square lattice, separate averages are performed in the two sublattices. The presence of a quantum critical point at $T = 0$ produces a fan of quantum fluctuations that can still be perceived at low enough temperatures\;\cite{Sachdev_2011}, suggesting that obtaining a signature of a quantum phase transition as a proxy at finite temperatures of $S$ might be possible.


\subsection{Systems}

We calculate the average single-site entanglement $S$ for the Hubbard model in honeycomb and kagome lattices and for the ionic Hubbard model in the square lattice. 
These systems have been previously studied and present correlation-driven quantum phase transitions.

The repulsive Hubbard model in the honeycomb lattice [Fig. \ref{fig:lattices}(a)] is known to display a quantum phase transition from a paramagnetic semimetal to an antiferromagnetic Mott insulator at half-filling\;\cite{Paiva_2005, Assaad2013, Otsuka2016} at $U_c/t \approx 3.8$. This bipartite geometry allows the possibility of particle-hole symmetry (PHS), which ensures that the behavior of quantities that depend on the electronic density is mirrored around the half-filling. Thus, it is sufficient to restrict the analysis to either above or below this point. We perform DQMC simulations on a honeycomb lattice of linear size $L = 9$ (N = $2 \times 9 \times 9$ sites) unless otherwise specified.

The absence of bipartite symmetry in the kagome lattice [Fig.\;\ref{fig:lattices}(b)] leads to a decoupling between the onset of insulating behavior and magnetic ordering with increasing interactions. Nonetheless, the existence of an MIT in the kagome Hubbard model has been extensively investigated. While earlier studies with different biased methodologies suggested that the MIT should occur at $U_c/t \in [5, 11]$\;\cite{ohashi_2006,Kuratani_2007,Yamada_2011,Higa_2016}, recent unbiased DQMC investigations have narrowed down where this transition should occur. In particular, Ref.\;[\onlinecite{kaufmann_2021}] combined dynamical mean-field theory, dynamical vertex approximation, and DQMC calculations, to find a transition at $U_c/t \in [7,9]$, while a recent work by some of us\;\cite{Medeiros-Silva2023} identified the transition at $U_c/t = 6.5 \pm 0.5$. Focusing on small-radius cylinder geometries, where the density-matrix renormalization group (DMRG) method is especially suitable, characterization of the onset of insulating behavior was established at $U_c/t \simeq 5.4$\;\cite{sun_2021}. Owing to the absence of PHS for any filling in the kagome lattice, which, as mentioned in Sec.\;\ref{DQMC_method}, leads to the minus-sign problem and limits the temperatures that can be accessed within DQMC, in this work, we focus the DQMC simulations on a kagome lattice of size $L = 6$ ($N = 3 \times 6 \times 6$ sites). 



\begin{figure}
\includegraphics[width=0.9\columnwidth]{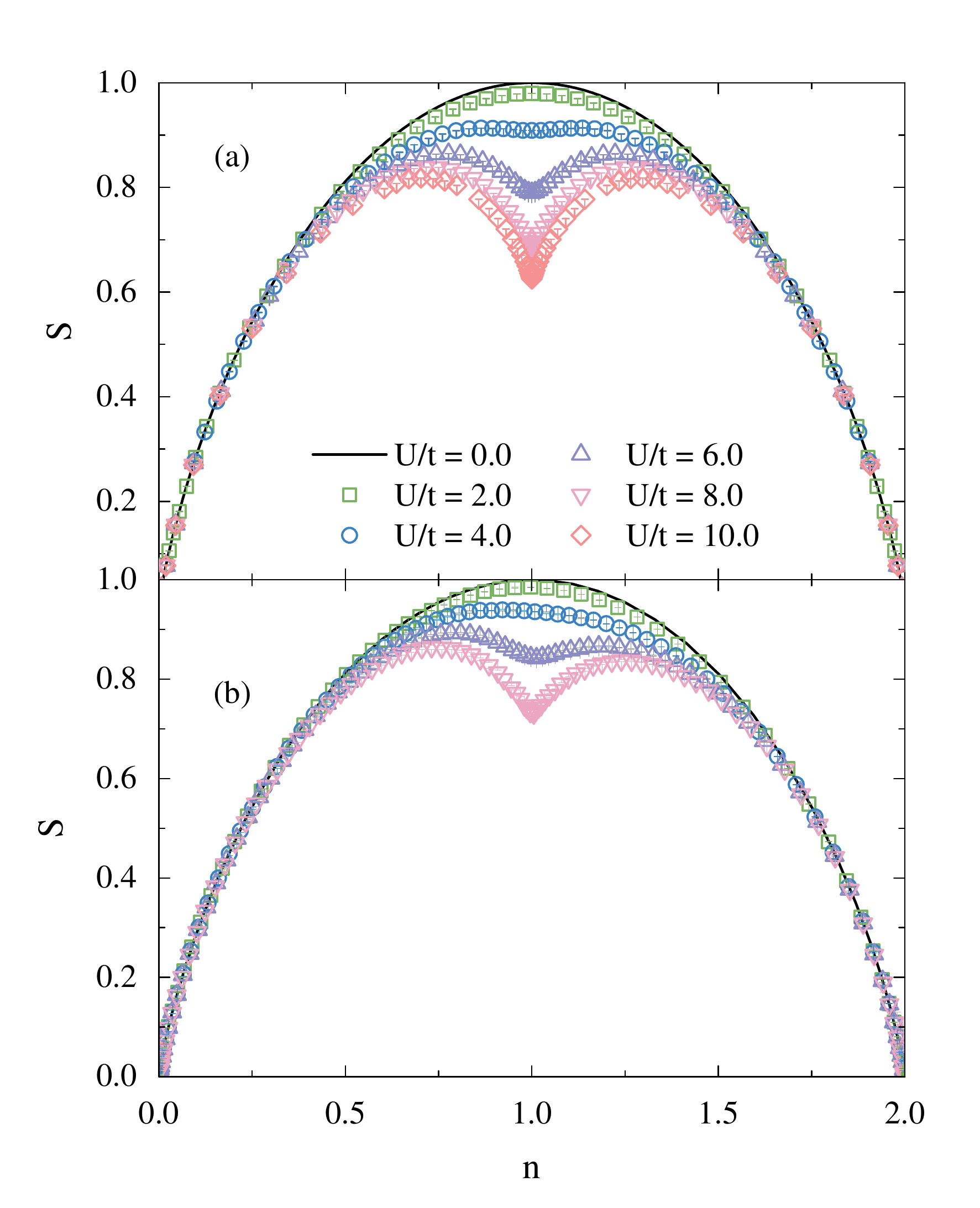}  
\caption{Average single-site entanglement $S$ as a function of density for different on-site interactions $U/t$ at $T/t=0.5$, for (a) honeycomb and (b) for kagome lattices. The lines give the non-interacting result at the same temperature.}
\label{S_n}
\end{figure}

The Hubbard model on a square lattice has a quantum phase transition at half-filling from a metallic paramagnet to an antiferromagnetic insulator as soon as the interactions are turned on. The transition is driven by the instability caused by the van Hove singularity and also by the presence of nesting\;\cite{Hirsch_1985,Hirsch_1987}. As finite-size effects are stronger for smaller $U$, and the gap is exponentially small in this regime, observing the transition through entanglement for the Hubbard model on a square lattice, where $U_c=0$, is challenging.

Therefore, to study the phase transition in the square lattice, we turn our attention to the ionic Hubbard model, where two phase transitions, driven by the on-site interaction $U$ at a finite $\Delta$\;\cite{Garg_2006,Bouadim_2007,Paris_2007,byczuk_2009} take place. The presence of $\Delta$ at weak interactions ($U/t \leq \Delta$) drives the system into a band insulating phase (BI). As $U$ increases, a metallic phase is found on a 2D square lattice\;\cite{Bouadim_2007,craco_2008}, whereas at large $U$ values, the system becomes a Mott insulator (MI)\;\cite{kancharla_2007}. Although the exact $U$-value where the Mott transition takes place is not well established, it is known that for $\Delta/t = 0.5$ at half-filling, the system undergoes a band-insulator-to-metal transition at $U/t \simeq 2$\;\cite{Bouadim_2007,Paris_2007,Mondaini_2023}. Thus, we focus on this value of staggered potential and perform DQMC simulations for a half-filled ionic square lattice [Fig.\;\ref{fig:lattices}(c)] of linear size $L = 10$ ($N = 10 \times 10$ sites).

In what follows, we calculate the single-site entanglement, in particular investigating how it varies with the electronic density $n$, in Sec.\;\ref{singlesite_densi}, and with the Hubbard interaction $U/t$, in Sec.\;\ref{singlesite_U}. The goal is to find regimes of parameters where indications of phase transitions emerge and compare the critical points obtained in previous studies. Thus, we aim to identify the reliability of the entanglement entropy calculated at $T > 0$ as a probe for quantum phase transitions in two dimensions.

\section{Results} \label{Results}

\subsection{Single-site entanglement as a function of  the electronic density} \label{singlesite_densi}


\begin{figure}[t]
\includegraphics[width=0.9\columnwidth]{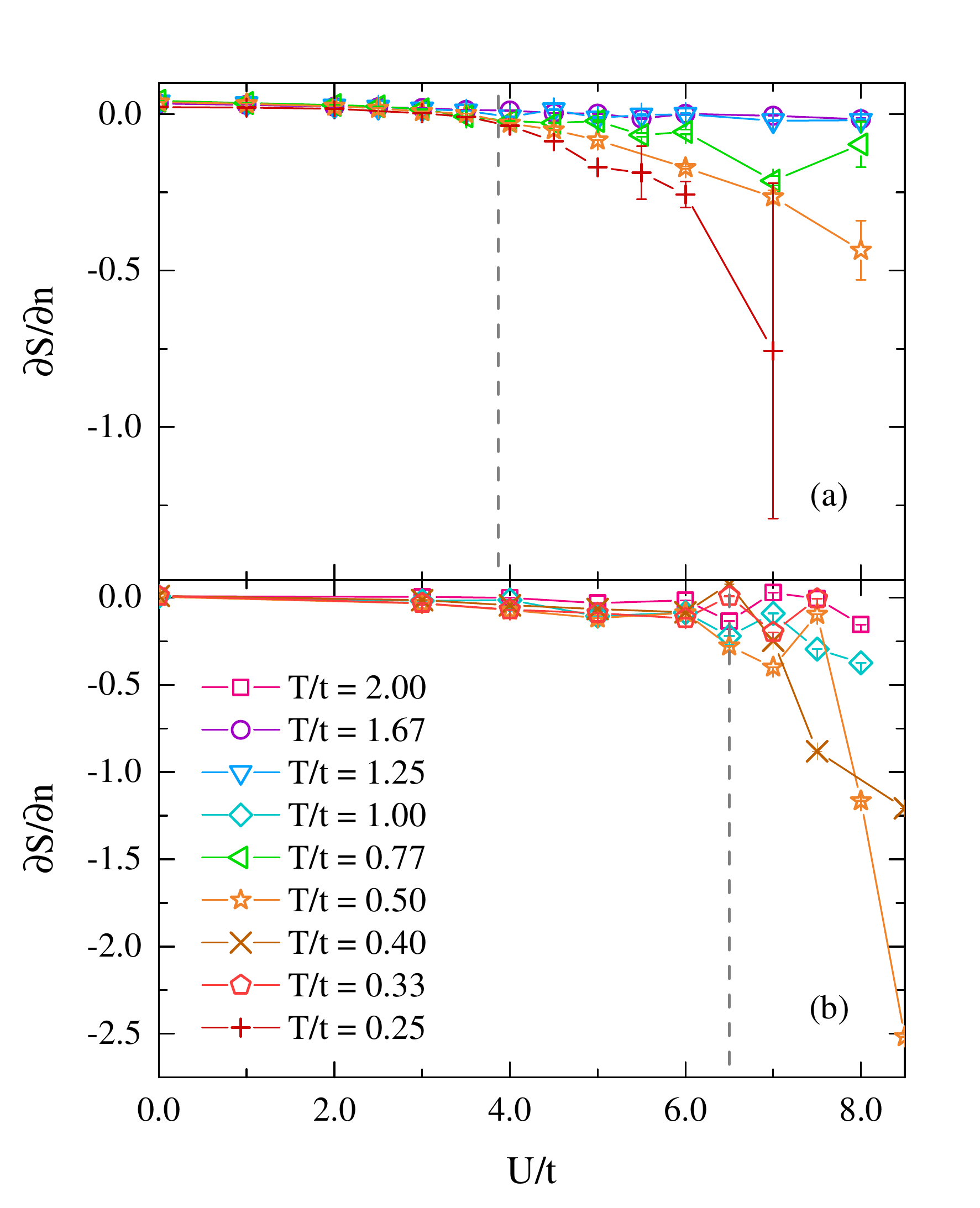} 
\caption{The $ U$-dependence of the derivative of the single-site entanglement with respect to the density at $n=1$ for (a) the honeycomb and (b) kagome lattices at different temperatures. Vertical dashed lines mark the critical interaction where $\partial S/\partial n$ significantly departs from zero at sufficiently low-$T$s, associated with the onset of insulating behavior -- see text. }
\label{fig:S_and_dS_dn_vs_U}
\end{figure}

We start by investigating the dependence of the entanglement entropy $S$ with the electronic density $n$ at $T/t = 0.5$, in Fig.\;\ref{S_n}.
The non-interacting limit shows a maximum of $S = 1$ at $n = 1$, consistent with the expectation that, at this regime, the system is metallic and each state has equal probability, i.e., $\omega_i=1/4$, leading to maximum entanglement. As $U/t$ increases, the entanglement starts to decrease, and the curve eventually changes into two maxima with a dip at $n = 1$. The appearance of a dip at half-filling signals a transition from the physics of itinerant electrons to that of localized moments, and it is interpreted as a marker of a Mott phase with increasing $U/t$\;\cite{Franca_2006, Arisa_2020}. As $U/t$ increases further, $U\rightarrow\infty$, the average single-site entanglement reaches the minimum $S=0.5$ at $n=1$, as charge degrees of freedom are frozen and only spin degrees of freedom remain\;\cite{Canella_2021}. It is worth pointing out that the two resulting domes in the curves for high $U/t$ are symmetric around $n=1$ for the honeycomb lattice, but not for the kagome lattice, due to the presence or absence of PHS in these systems.

\begin{figure*}[t]
\includegraphics[width=0.9\textwidth]{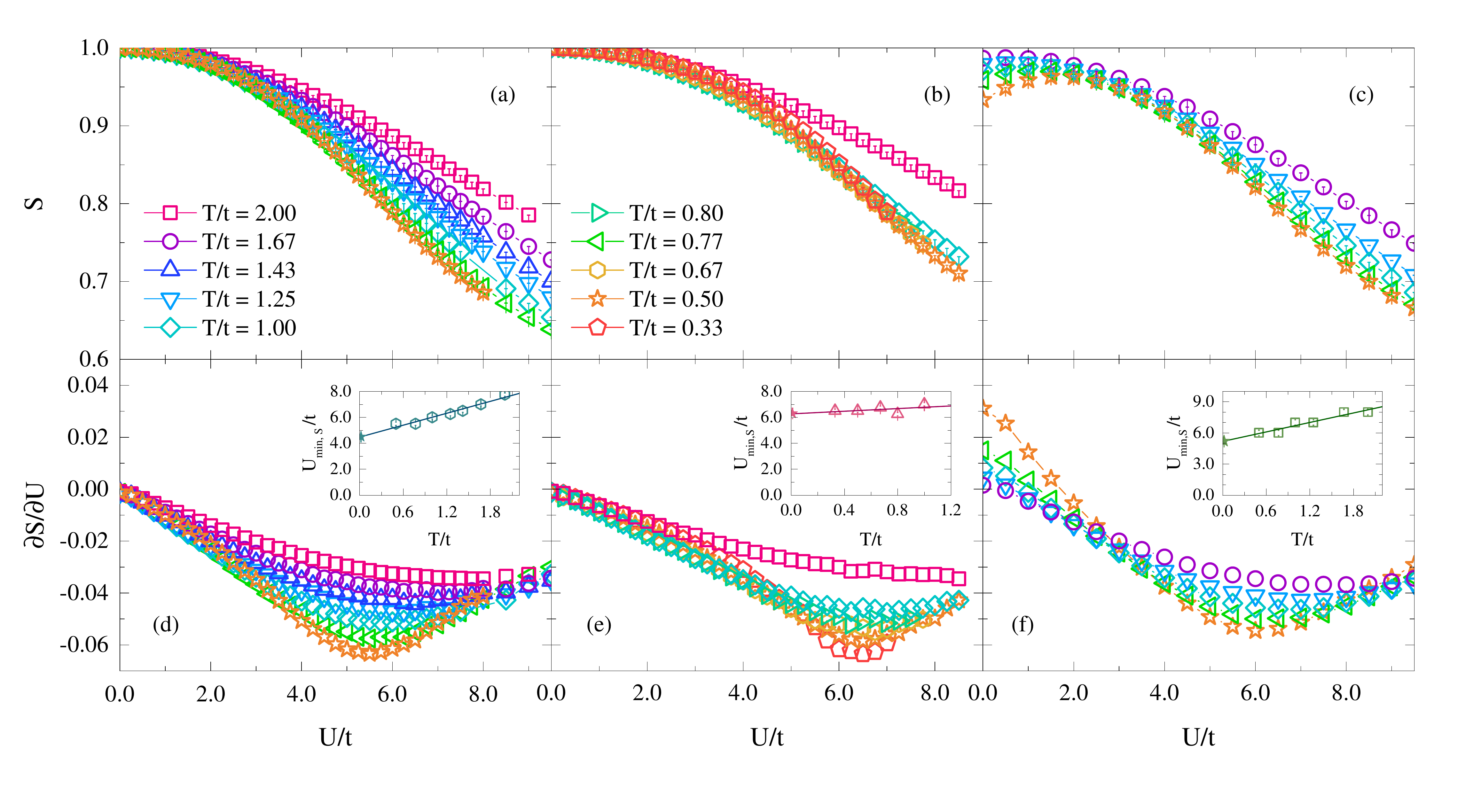} 
\vspace{-0.1in}
\caption{Average single-site entanglement $S$ as a function of on-site interaction $U/t$ for (a) honeycomb, (b) kagome, and (c) ionic square lattices, contrasting different temperatures at half-filling ($n=1$). The decrease at large $U$ reflects the localizing tendency of the electrons in this regime. Panels (d), (e), and (f) show the corresponding derivatives with respect to the interaction strength. The minima signal a finite-$T$ proxy of the onset of insulating behavior -- its temperature dependence is shown in the insets. The lines correspond to linear fits from which the critical values at $T=0$ ($U_c/t$) are extracted, resulting in $U_c/t = 4.5(2)$ for the honeycomb lattice, $U_c/t = 6.3(4)$ for the kagome lattice, and $U_c/t = 5.2(3)$ for the ionic square lattice, depicted by the star markers in the corresponding insets.}  
\label{fig:S_and_dS_dU_vs_U}
\end{figure*}

To monitor the temperature effects in our results, we investigate the derivative $\partial S/\partial n$ around half-filling for different $T$'s as shown in Fig.\;\ref{fig:S_and_dS_dn_vs_U}. For lower values of $T/t$, $\partial S/\partial n = 0$ for $U < U_c$ whereas $\partial S/\partial n \neq 0$ for $U > U_c$.
This is consistent with the fact that in a metallic phase, there are no significant changes in the non-local correlations when adding/subtracting a particle to/from the system, while in the insulating phase, quantum correlations become more susceptible to the addition/subtraction of particles\;\cite{Gu_2004}.
This change in regimes happens when $U/t$ is increased beyond the critical values $U_c$,
indicating that $\partial S/\partial n$ can signal the MIT.
However, when $T/t$ increases, the entanglement entropy shows a less pronounced response, suggesting that for $T/t \gtrsim 1$ the thermal fluctuations are intense enough to wash away the effects of the quantum phase transition.

\subsection{Single-site entanglement as a function of interaction strength} \label{singlesite_U}
\begin{figure}[t]
\includegraphics[width=0.9\columnwidth]{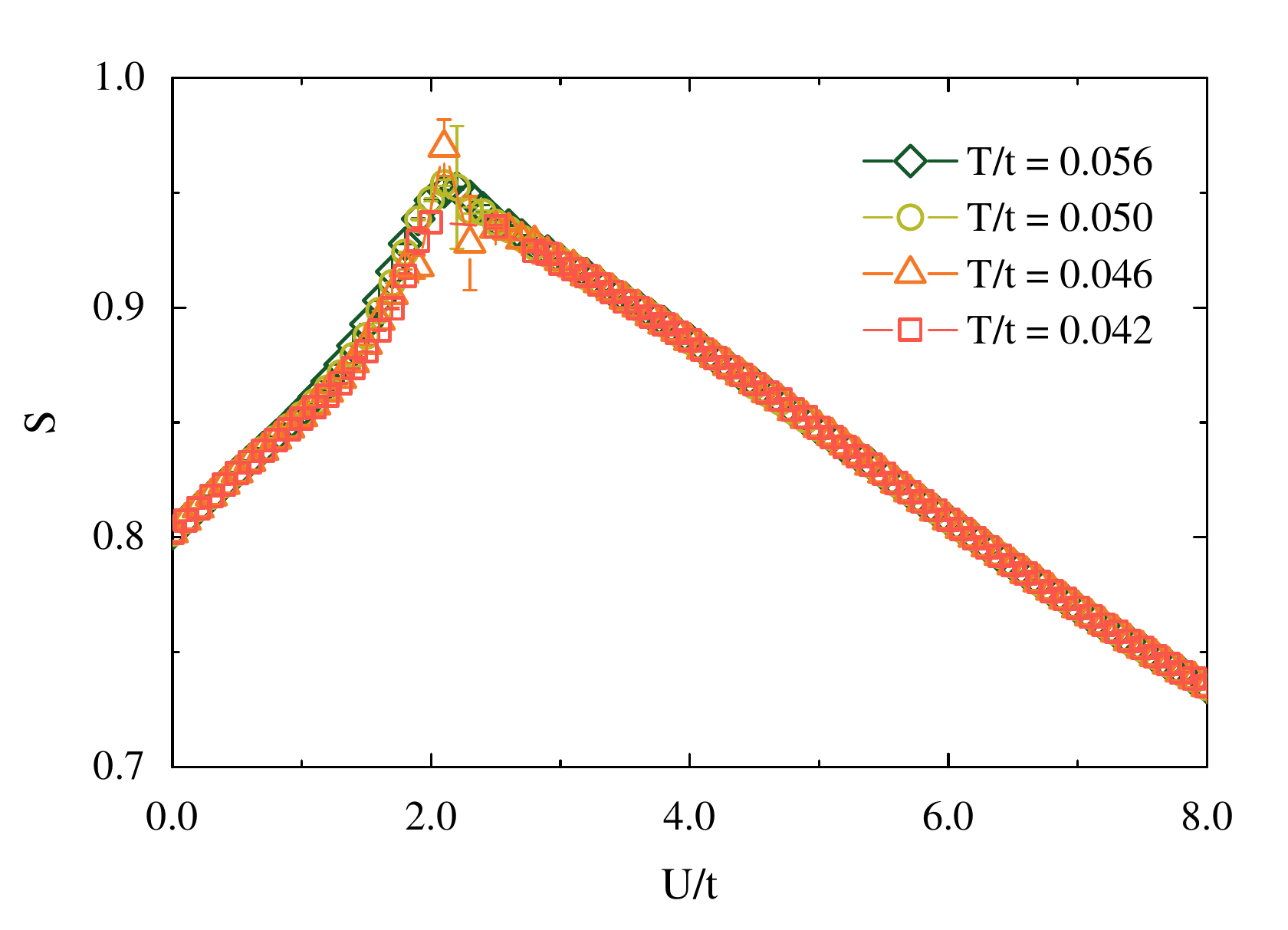} 
\caption{Single-site entanglement as a function of $U/t$ for the ionic square lattice for a  $ 4 \times 4$ lattice and a temperature range $ 0.04 < T/t < 0.06$ -- the peak at $U/t \simeq 2$, sharper at lower $T$s, gives the location of the band-insulator-to-correlated-metal transition.}
\label{SVN_ioni_N4_betas}
\end{figure}

Focusing on half-filling where the Mott transition occurs, we now investigate the Hubbard interaction $U$ effects on $S$. Figure\;\ref{fig:S_and_dS_dU_vs_U} shows that the honeycomb [Fig.\;\ref{fig:S_and_dS_dU_vs_U}(a)] and kagome lattices [Fig.\;\ref{fig:S_and_dS_dU_vs_U}(b)] display similar behavior, where the entanglement is maximum in the non-interacting regime while decaying monotonically with increasing $U/t$. This happens as a consequence of the zero and double occupancies, $\omega_{i,0}$ and $\omega_{i,\uparrow \downarrow}$, decreasing while the single occupancies, $\omega_{i,\sigma}$, increase, thus indicating a localizing tendency of the electrons. In contrast, $S$ displays a non-monotonic behavior for the ionic Hubbard model, increasing with $U$ until $U/t \simeq 2$ and then decreasing as the Hubbard correlation further increases -- it directly derives from a competition between $\Delta$ and $U$. In particular, this model presents a phase transition from a band insulator to a correlated metal for $\Delta/t = 0.5$ at $U/t \simeq 2$ \;\cite{craco_2008,Mondaini_2023}, in agreement with the maxima observed at low temperatures in Fig.\;\ref{fig:S_and_dS_dU_vs_U}(c).  

As this peak is not present for the other cases examined, we further explore its behavior at $T \simeq t/20 $, shown in Fig.\;\ref{SVN_ioni_N4_betas} for a $4 \times 4$ lattice. The peak turns sharper at this low-temperature limit, indicating that the band-insulator-to-metal transition is signaled by the single-site von Neumann entropy rather than its derivative. Notably, in such a small lattice size, it has been noted that this is a first-order phase transition, i.e., associated with energy level crossing\;\cite{mondaini2022}. The metallic state survives the increased interactions until the screened Coulomb repulsion leads to the gap reopening, when a second phase transition is observed, from a correlated metal to a Mott insulator.


To explore the quantitative description of the onset of the Mott regime, we analyze the behavior of $\partial S/\partial U$ in Fig.\;\ref{fig:S_and_dS_dU_vs_U} for the different lattices. 
For the honeycomb [Fig.\;\ref{fig:S_and_dS_dU_vs_U}\;(d)] and ionic square lattices [Fig.\;\ref{fig:S_and_dS_dU_vs_U}\;(f)], we verify that $\partial S/\partial U$ present minima at $U_{\rm min, S}(T)$ that slowly move towards smaller values of $U/t$ as the temperature decreases. On the other hand, for the kagome lattice [Fig.\;\ref{fig:S_and_dS_dU_vs_U}\;(e)], all minima are located at similar values of $U/t$ for all $T/t$. To obtain an estimate for the critical value of the interaction strength $U_c$ at which the Mott transition takes place, we extrapolate the minima positions to $T\rightarrow 0$, assuming a linear dependence -- see insets in Fig.\;\ref{fig:S_and_dS_dU_vs_U}. Notice that the minima in $\partial S/\partial U$ are better defined for the lower temperatures examined and flatten as the temperature increases, disappearing for $T/t \gtrsim 2$.

\begin{figure}[th]
\includegraphics[width=0.9\columnwidth]{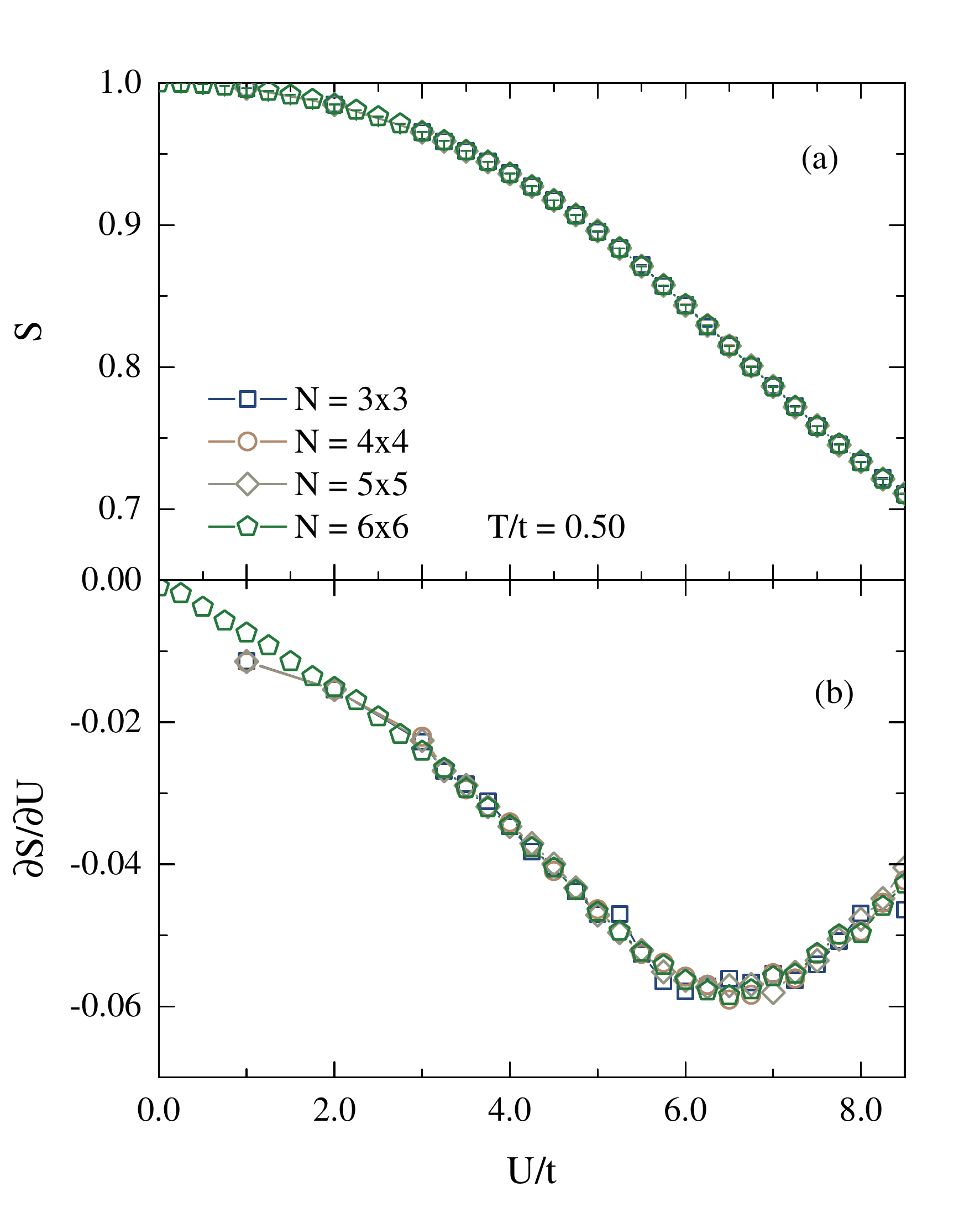} 
\caption{Average single-site entanglement (a) and its derivative with respect to $U/t$ (b)  as a function of $U/t$ for different system sizes for the kagome lattice at $T/t = 0.5$. 
}
\label{fig:kg_N3_N6_9}
\end{figure}

Remarkably, the critical values obtained from the $T\to 0$ extrapolation are in agreement to those previously found for the Mott transition for all three lattices\;\cite{Paiva_2005,Assaad2013,Otsuka2016,Medeiros-Silva2023,craco_2008,Mondaini_2023}, confirming that the average single-site entanglement is a good indicator of quantum phase transitions even at finite temperature for $T/t \lesssim 1$. Since these results are extracted at fixed system sizes, we now probe the potential influence of finite-size effects. Indeed, focusing on the kagome lattice, Fig.\;\ref{fig:kg_N3_N6_9} shows that the single-site entanglement and its derivative at $T/t=0.5$ exhibit quantitatively minor finite-size effects for the analyzed range of number of sites, thus showing the robustness of the computed values of $U_c$ to trigger the MI regime.

\subsection{Susceptibility and entanglement}


In this section, we analyze the relation between the single-site entanglement $S$ and the ferromagnetic spin susceptibility  $\chi(q=0)$, obtained through the relation\;\cite{Hirsch_1985}
\begin{equation}
    \label{suscep}
    \chi(q=0) = - \beta \;\frac{1}{N} \sum_{\substack{ \textbf{i},\textbf{j} }} \langle (\hat n_{{\bf i}\uparrow} - \hat n_{{\bf i} \downarrow})(\hat n_{{\bf j} \uparrow} - \hat n_{{\bf j} \downarrow}) \rangle\ .
\end{equation}

\begin{figure}[t]
\includegraphics[width=0.9\columnwidth]{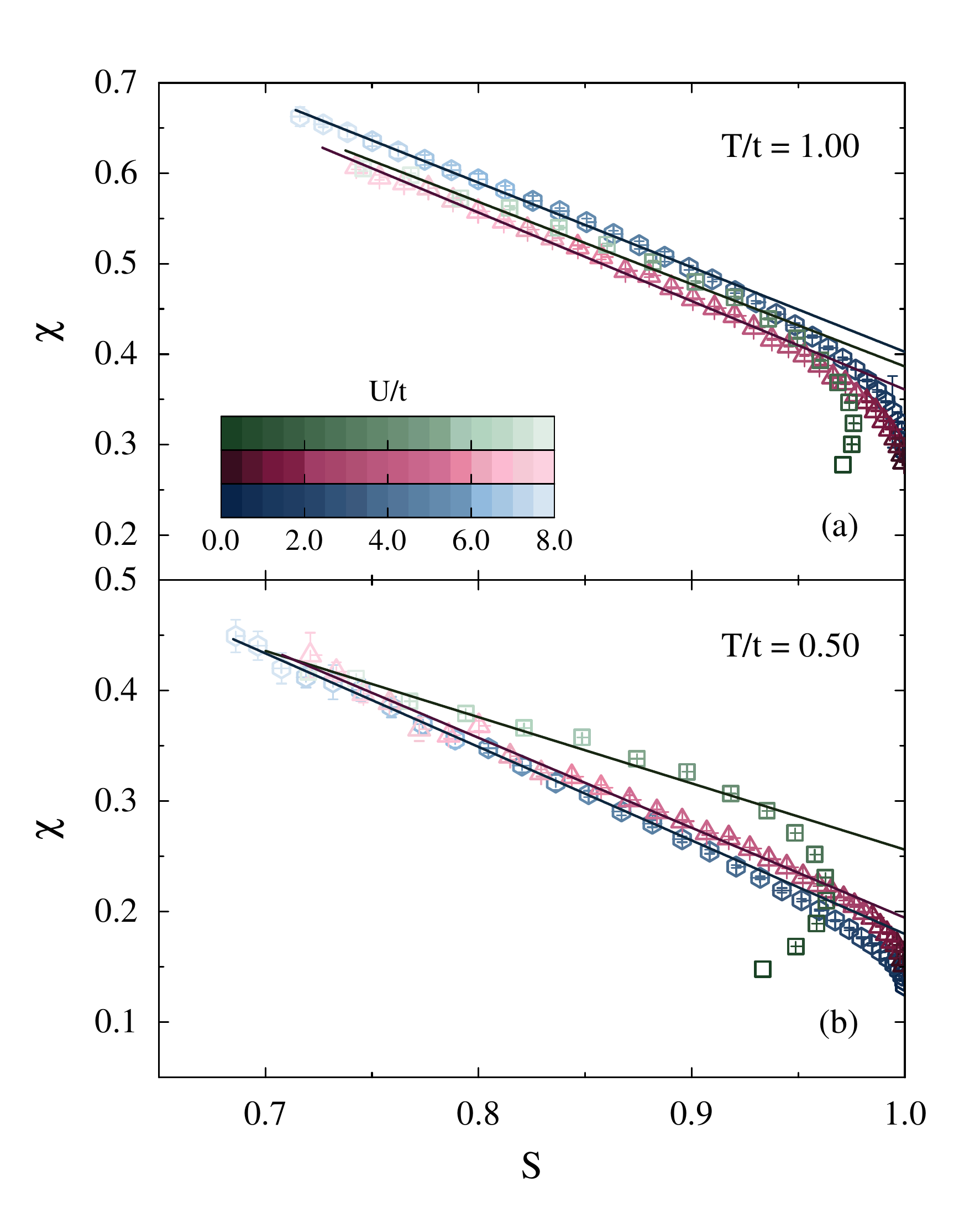}
\caption{Ferromagnetic susceptibility as a function of single-site entanglement for the honeycomb (blue hexagons), kagome (pink triangles), and ionic (green squares) lattices for $T/t = 1.00$\;(a) and $T/t = 0.50$\;(b). The dark-colored lines correspond to linear fits.}
\label{fig:en_chi_vs_S}
\end{figure}

Focusing on half-filling, we report the $\chi$ vs.\;$S$ relation in Fig.\;\ref{fig:en_chi_vs_S} for the honeycomb, kagome and ionic square lattices, at $T/t = 0.50$ and $T/t = 1.00$. The on-site interaction parametrizes the plot and increases from right to left (see color bars).
While the single-site entanglement decreases as $U/t$ increases (which was expected due to the behavior observed in Figs.\;\ref{S_n} and \ref{fig:S_and_dS_dU_vs_U}), the susceptibility increases with $U/t$.
These might be connected with the enhancement of local moment formation and short-range spin-spin correlation functions which are a consequence of the decrease of empty and double occupancies ($\omega_{i,0}$ and $\omega_{i,\uparrow\downarrow}$, respectively)\;\cite{Hirsch_1985,moreo_1993}.

\begin{figure}[t]
\includegraphics[width=0.9\columnwidth]{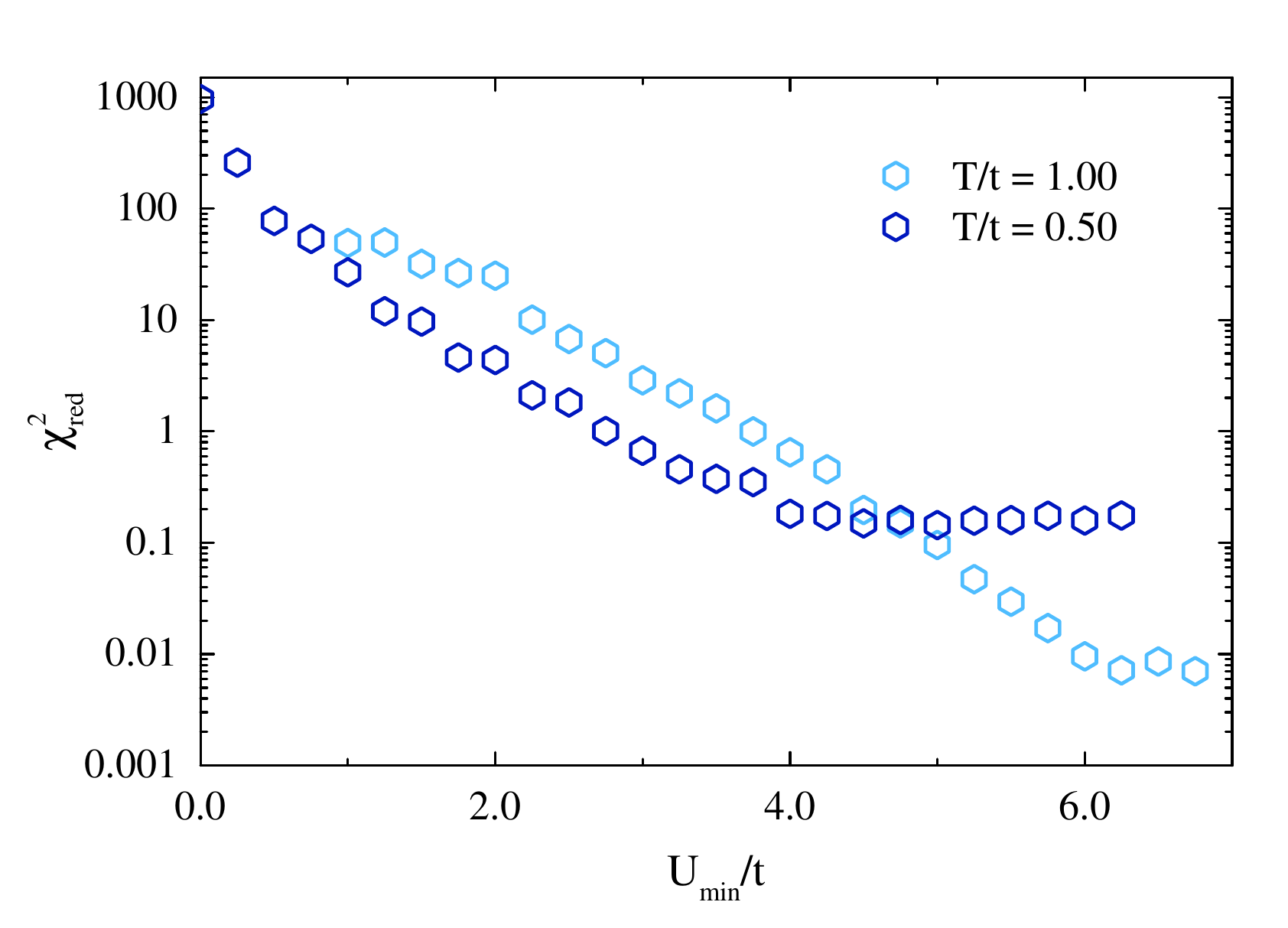} \caption{Reduced $\chi^2$, $\chi^2_{\rm red}$, for the linear fits of the susceptibility for the honeycomb lattice featured in Fig.\;\ref{fig:en_chi_vs_S}. Each point is obtained by performing a linear fit within the $[U_{\rm min},U_{\rm max}]$ interval -- see text.}
\label{red_chi_hon}
\end{figure}

An interesting feature in Fig.\;\ref{fig:en_chi_vs_S} is the linear dependence between the susceptibility and $S$, similar to that observed in one-dimensional superfluid Hubbard chains \cite{Arisa_2020}. The linear behavior is well established for higher values of $\chi$, with the maximal value of $\chi_{\rm max}$ being equivalent to $U_{\rm{max}}/t = 8.0$ (lightest points in Fig.\;\ref{fig:en_chi_vs_S}), and starts from a certain $\chi_{\rm min}$, with corresponding $U_{\rm min}/t$. To understand the relation of this linear behavior with $U/t$, we performed linear fits where we set the maximal value to be $\chi_{\rm max}$ and varied $\chi_{\rm min}$ from lowest to highest, thus obtaining linear fits for the interval $[U_{\rm min}/t, U_{\rm{max}}/t]$, with varying $U_{\rm min}/t$. From these fits, we obtained the reduced $\chi^2$, $\chi_{\rm red}^2 = \chi^2 / \rm{degrees \; of \; freedom}$, which we plotted against $U_{\rm min}/t$ in Fig.\;\ref{red_chi_hon}. For the honeycomb lattice at $T/t=0.50$, $\chi_{\rm red}^2$ decreases as $U_{\rm min}/t$ increases for $U_{\rm min}/t < 4$, developing a plateau for $U_{\rm min}/t \geq 4$. This change of behavior happens near the critical value for the honeycomb lattice, which may suggest a connection between the linear behavior of the parametrized susceptibility and the Mott transition. For $T/t=1.00$, however, the plateau only develops for $U_{\rm min}/t \geq 6$. The dependence of the linear behavior of $\chi$ on the temperature is similar to that of $U_{\rm min,S}$ in Fig.\;\ref{fig:S_and_dS_dU_vs_U}\;(d).

\section{Conclusions} \label{conclusions}

We study the behavior of the single-site entanglement entropy in two-dimensional systems at finite temperatures. Our goal was to investigate whether the von Neumann entropy at $T \neq 0$ can still define bipartite entanglement and/or be used to identify and characterize Mott metal-insulator transitions in two-dimensional lattices. To this end, we explore the relation between the single-site entanglement entropy and the critical interaction strengths for the onset of a Mott Insulator in the Hubbard model on the honeycomb and kagome lattices. For the ionic Hubbard model on the square lattice, we found that $S$ signals both the Mott and the band insulator to metal transition.

When analyzing the dependency of the single-site entanglement with the electronic density, we see a change in the behavior of $S$, going from a one- to a two-peaked structure with increasing $U/t$ at half-filling. 
This behavior change can be observed for temperatures $T/t  \leq 1.00$, and is caused by fluctuations produced by the quantum phase points at $T = 0$.

While the dependency of the single-site entanglement with the onsite interaction does not provide enough information to identify a Mott transition at finite temperatures for the Hubbard model, it does signal a metal-to-band-insulator transition for the ionic Hubbard model on the square lattice, providing a critical value $U_c$ in accordance with previous results\;\cite{craco_2008,Mondaini_2023}.
The derivative $\partial S/ \partial U$ presents minima for all systems analyzed at temperatures $T/t \lesssim 1.00$, which, when extrapolated to the $T \to 0$ limit, allow us to find values of $U_c/t$ that agree with previous works.
Combined with the previous findings, these results show that single-site entanglement can be employed to estimate quantum phase transitions at finite temperatures.

We also examined the connection between entanglement and magnetic susceptibility. 
We found that the parametrized magnetic susceptibility has a linear dependence on the single-site entanglement for all systems, similar to one-dimensional Hubbard chains.
While a linear relation between the magnetic susceptibility and the entanglement is insufficient to quantitatively determine the latter, it can still be used to qualitatively estimate the degree of entanglement or to compare the entanglement between different systems from an experimental perspective.

\section*{ACKNOWLEDGMENTS}
The authors are grateful to the Brazilian agencies Conselho Nacional de Desenvolvimento Cient\'\i fico e Tecnol\'ogico (CNPq) and Coordena\c c\~ao de Aperfei\c coamento de Pessoal de Ensino Superior (CAPES).
We acknowledge financial support from Funda\c c\~ao  Carlos Chagas Filho de Amparo \`a Pesquisa do Estado do Rio de Janeiro, grant numbers E-26/204.308/2021 (W.C.F.S.),  E-26/200.959/2022 (T.P.), E-26/210.100/2023 (T.P.); S\~ao Paulo Research Foundation (FAPESP), grant number 2021/06744-8 (V.V.F.); CNPq, grant numbers 403890/2021-7 (V.V.F.), 306301/2022-9 (V.V.F.), 308335/2019-8 (T.P.) 403130/2021-2 (T.P.), 442072/2023-6 (T.P.); and also Instituto Nacional de Ci\^encia e Tecnologia de Informa\c c\~ao Qu\^antica (INCT-IQ). R.M.~acknowledges support from the T$_c$SUH Welch Professorship Award.


\bibliography{Entanglement.bib}

\end{document}